\documentclass[manuscript, screen]{acmart}

\usepackage{listings}
\usepackage{makecell}
\usepackage{textcomp}
\usepackage{enumitem}
\usepackage{balance}
\usepackage{soul}
\usepackage{xcolor}
\usepackage{colortbl}
\definecolor{darkgreen}{RGB}{0, 205, 130}
\definecolor{light-red}{RGB}{250, 160, 160}
\definecolor{light-green}{RGB}{160, 250, 160}
\definecolor{my-gray}{RGB}{100, 100, 100}
\definecolor{light-blue}{RGB}{150, 200, 250}
\definecolor{very-light-blue}{RGB}{215, 230, 250}
\definecolor{etonblue}{rgb}{0.60, 0.89, 0.65}
\definecolor{light-gray}{gray}{0.97}
\definecolor{just-gray}{gray}{0.90}
\sethlcolor{green}

\copyrightyear{2024} 
\acmYear{2024} 
\setcopyright{acmlicensed}\acmConference[Koli Calling '24]{24th Koli Calling International Conference on Computing Education Research}{November 12--17, 2024}{Koli, Finland}
\acmBooktitle{24th Koli Calling International Conference on Computing Education Research (Koli Calling '24), November 12--17, 2024, Koli, Finland}
\acmDOI{10.1145/3699538.3699550}
\acmISBN{979-8-4007-1038-4/24/11}

\begin{document}

\title{Investigating Student Reasoning in Method-Level Code Refactoring: A Think-Aloud Study}

\author{Eduardo Carneiro Oliveira}
\affiliation{%
   \institution{Utrecht University}
   \country{The Netherlands}}
\email{e.carneirodeoliveira@uu.nl}

\author{Hieke Keuning}
\affiliation{%
   \institution{Utrecht University}
   \country{The Netherlands}}
\email{h.w.keuning@uu.nl}

\author{Johan Jeuring}
\affiliation{%
   \institution{Utrecht University}
   \country{The Netherlands}}
\email{j.t.jeuring@uu.nl}

\begin{abstract}
Producing code of good quality is an essential skill in software development. Code quality is an aspect of software quality that concerns the directly observable properties of code, such as decomposition, modularization, and code flow. Code quality can often be improved by means of code refactoring -- an internal change made to code that does not alter its observable behavior. According to the ACM/IEEE-CS/AAAI Computer Science Curricula 2023, code refactoring and code quality are core topics in software engineering education. However, studies show that students often produce code with persistent quality issues. Therefore, it is important to understand what problems students experience when trying to identify and fix code quality issues. In a prior study, we identified a number of student misconceptions in method-level code refactoring. In this paper, we present the findings from a think-aloud study conducted to investigate what students think when working on method-level refactoring exercises. We use grounded theory to identify and classify students’ reasonings. As a result of the analysis, we identify a set of eight reasons given by students to refactor code, which either concerns the presence of code quality issues, the improvement of software quality attributes, or code semantics. We also analyze which quality issues are identified by students, and to which reasonings these quality issues are related. We found that experienced students reason more often about code quality attributes rather than pointing at a problem they see in the code. Students were able to remove code quality issues in most cases. However, they often overlooked particular code quality issues, such as the presence of a method with multiple responsibilities or the use of a less suitable loop structure.
\end{abstract}

\ccsdesc[500]{Social and professional topics~Computer science education}
\ccsdesc[500]{Social and professional topics~Software engineering education}

\keywords{code refactoring; code quality; student reasoning; think-aloud; program snapshot analysis; grounded theory; programming education}

\maketitle

\section{Introduction}
\label{Introduction}

Producing code of good quality is an essential skill in software development. \textit{Code quality} can be defined as an aspect of software quality that concerns the directly observable properties of code, such as decomposition, modularization and flow \cite{stegeman2016designing}. Writing programs that adhere to quality standards may help developers read, maintain, and fix code \cite{fowler2018refactoring}. Code quality can often be improved by means of \textit{refactoring} \cite{fowler2018refactoring}. Code refactoring is an internal change made to code that does not alter its observable behavior. Determining the quality of code may become even more important with the increasing use of AI coding assistants, since automatically generated code often needs to be refactored \cite{yeticstiren2023evaluating, gitclear2024}.

In the ACM/IEEE-CS/AAAI Computer Science Curricula 2023 \cite{curricula2023}, code refactoring is presented as a core section in Software Engineering, and code quality appears as a core topic in the software design section. Therefore, we expect it will be taught in quite a few CS programs. However, studies analyzing the quality of code developed by students found that students often write code with quality issues, and that some of these issues are never fixed \cite{edwards2017investigating, effenberger2022code, ostlund2023s, tigina2023analyzing, keuning2017code}. Hence, it is important to look at what problems students experience when learning about identifying and removing code quality issues. 
Such issues appear both in small programs novices write, as well as in larger programs written by more advanced students. In this study, we focus on small programs written by novices. For them, relevant refactorings are typically at the method or statement-levels. In a previous study \cite{oliveira2023student}, we identified a number of common misconceptions that students hold when refactoring method-level code. This paper extends that study by investigating how students \textit{reason} when they work on method-level code refactoring exercises. Examples of \textit{reasoning} may include, but are not limited to, trying to understand the code purpose, choosing to perform a refactoring, or identifying a quality issue. These reasonings may offer an explanation for some of the refactoring misconceptions, and may help explain why and how students identify and remove code quality issues. We address the following research questions:

\vspace{1mm}

\textit{\textbf{RQ1.} What are students' reasonings when refactoring code?}


\textit{\textbf{RQ2.} What is the relation between the code quality issues addressed by students and their reasonings?}


\vspace{1mm}

The main contributions of our study are: (1) a categorization of students' reasonings when working on method-level code refactoring exercises; and (2) an analysis of the relation between students' reasonings and the code quality issues addressed by them. 


\section{Background and Related Work}
\label{Background}

This section describes terms and definitions relevant for our study, as well as related studies on code quality in education, refactoring misconceptions and student reasoning.

\subsection{Terms and definitions}
\label{Definitions}

\textit{Code quality.} \citeauthor{stegeman2016designing} \cite{stegeman2016designing} define \textit{code quality} as an aspect of software quality concerning properties of code that can be directly observed. They propose a rubric that comprises a set of criteria such as code flow, decomposition and modularization. \citeauthor{kirk2024literature} \cite{kirk2024literature} use the term \textit{code style} to refer to code quality, and extend \citeauthor{stegeman2016designing}'s definition by adding that the criteria from their rubric are constrained to understanding and changing code.

\vspace{1mm}


\textit{Code quality attribute.} The ISO/IEC 25010 standard \cite{ISO25010} is a quality model with characteristics for evaluating properties of software products. In this paper, we use the term \textit{quality attribute} to refer to those characteristics. Examples of quality attributes are maintainability, performance and flexibility.

\vspace{1mm}

\textit{Code quality issue.} We use the term \textit{code quality issue} to describe a problem in code that makes one or more quality attributes worse. For example, code with excessive nesting has lower readability. Other terms used to refer to code quality issues are \textit{code smells} \cite{beck2018refactoring, mcconnell2004code}, \textit{antipatterns} \cite{nurollahian2023use}, and \textit{code quality defects} \cite{effenberger2022code, vrechtavckova2024catalog}.

\vspace{1mm}

\textit{Code refactoring.} \citeauthor{fowler2018refactoring} \cite{fowler2018refactoring} defines \textit{refactoring} as a change made to the internal structure of a program that does not modify its observable behavior. Examples of refactoring are extracting a method or simplifying conditional expressions. Code refactoring may be helpful for various reasons~\cite{fowler2018refactoring}. When multiple programmers with different goals change code over a period of time, code may lose its structure. By refactoring a fragment of code that is not ideally structured, it becomes easier to understand and maintain.

\subsection{Code quality in education}


A systematic mapping study \cite{keuning2023systematic} identifies 195 papers concerning code quality in education. This study shows that the main research focus has been on developing and evaluating tools for feedback on code smells, as well as suggestions for improvements and refactorings. It suggests that other topics should be further explored, such as student perceptions of code quality. \citeauthor{borstler2018know} \cite{borstler2018know} address this topic by investigating how students, educators and software developers perceive code quality. A few terms are often used among all groups to describe code quality, such as readability and structure. \citeauthor{wiese2019linking} \cite{wiese2019linking} examine students' perceptions of readability in code written by novices and experts. They find that a number of students consider code written by experts less readable than novice code. \citeauthor{da2023test} \cite{da2023test, silva2023misconceptions} present a set of code quality issues identified in CS1 student programs, along with an analysis of students' perceptions on the issues in their code. Students mention the use of redundant checks as a way to guarantee that their code passes all test cases. In addition, a few students demonstrated an incomplete comprehension of control structures, such as the use of iteration variables in for loops. 

Several studies focus on identifying quality issues in student code. \citeauthor{tigina2023analyzing} \cite{tigina2023analyzing} analyze the prevalence of code quality issues in student submissions, and how students attempt to fix particular issues indicated by a code quality assessment tool. They find that students frequently fix issues that require simple refactorings, such as removing unused code, but occasionally ignore issues that they find difficult to understand. Likewise, \citeauthor{keuning2017code} \cite{keuning2017code} examine quality issues in student submissions, and find that issues related to modularization are rarely fixed. \citeauthor{effenberger2022code} \cite{effenberger2022code} present a catalog of code quality issues (called \textit{defects}) found in student programs, and describe rules on how particular issues can be detected in code. \citeauthor{vrechtavckova2024catalog} \cite{vrechtavckova2024catalog} extend their work with a catalog of 80 defects. They propose a classification of defects into eight different types, such as \textit{poor design} or \textit{simplifiable code}. \citeauthor{nurollahian2023use} \cite{nurollahian2023use} investigate students' choices regarding the use of sequential if statements. They find several cases of a student using both desired patterns and anti-patterns in their solutions.
Other studies also analyze the presence of quality issues in student code \cite{breuker2011measuring, de2018understanding}, or whether students improve code quality in their submissions during semester-long programming courses \cite{edwards2017investigating, ostlund2023s}. Prior studies give little insight into the reasons students give for choosing code structures or when to remove code quality issues. \citeauthor{nurollahian2023use} \cite{nurollahian2023use} advocate for think-aloud studies to examine the reasons behind student use of anti-patterns. In this paper, we investigate what students think while working on code refactoring tasks.

\subsection{Refactoring misconceptions}
\label{background-rm}

In a prior study \cite{oliveira2023student}, we define a \textit{refactoring misconception} as an incorrect refactoring step that results in semantically incorrect code, showing an inadequate understanding of a particular programming concept. We analyze log data containing program snapshots of students working on method-level refactoring exercises in a tutoring system, and identified a total of 25 misconceptions. These refactoring misconceptions are classified according to their related code structure: arithmetic expressions, boolean expressions, conditional statements, loops, or flow-related commands. An example of such a misconception, called \textit{IfsMerged}, appears in the green frame below. In this misconception a student merges two nested conditions. The red frame shows how the student uses an {\&\&}-operator to combine the boolean expressions, but does not take the else part of the outer-if into account when introducing the new boolean expression, resulting into semantically incorrect code.

\vspace{2mm}

{\centering
\begin{minipage}{2.7cm}
\begin{lstlisting}[basicstyle={\footnotesize\ttfamily\linespread{1.0}\selectfont}, frame=single, rulecolor=\color{teal}]
if (posOnly) {
    if (value >= 0) {
        sum += value;
    }
}
else {
    sum += value;
}
\end{lstlisting}
\end{minipage}
\hspace{0.2cm}
\textrightarrow
\hspace{0.3cm}
\begin{minipage}{3.5cm}
\begin{lstlisting}[basicstyle={\footnotesize\ttfamily\linespread{1.0}\selectfont}, frame=single, rulecolor=\color{red}]
if (posOnly && value >= 0) {
    sum += value;
}
\end{lstlisting}
\end{minipage} \\[2mm]
\par}
\noindent
To also address changes in code quality, we slightly extend this definition in this study: a refactoring misconception is an incorrect refactoring step that transforms semantically correct code into incorrect code, \textit{or introduces a code quality issue} -- for example, to unnecessarily increase code nesting.

\subsection{Student reasoning}

Prior studies in Computing Education have investigated student reasoning while performing a task. 
\citeauthor{kennedy2018they} \cite{kennedy2018they} investigate how students reason when working on CS1 exercises involving troublesome concepts, such as the use of global variables, and the use of functions with no return. They identify students' misconceptions related to parameter passing by value versus passing by references.
\citeauthor{izu2023exploring} \cite{izu2023exploring} conduct a study in which students are asked to rank a set of small solutions according to their own perceptions of code quality, and to give explanations for their choices. Performance and structure are the most common code quality attributes considered by students when ranking solutions.
In this paper, we investigate student reasoning when working on refactoring exercises. Their reasoning might explain some of the refactoring misconceptions we found in our previous work. We follow \citeauthor{kennedy2018they} \cite{kennedy2018they} by using the term \textit{reasoning} to refer to students' thoughts and explanations for choices during working on a task. Similar to \citeauthor{izu2023exploring} \cite{izu2023exploring}, we are interested in students' perceptions of code quality in method-level programs. However, their study focuses on the reasons students give to rank solutions in terms of code quality. Our study focuses on student reasoning when working on method-level refactoring tasks, which is closer to the activities of software developers, and therefore more authentic.

\section{Method}
\label{Method}


\subsection{Think-Aloud Method}

Qualitative studies can be used to collect three basic types of phenomena: \textit{activities}, \textit{accounts} and \textit{artifacts}~\cite{tenenberg2019qualitative}. \textit{Activities} refer to what a participant does, such as writing a computer program; \textit{accounts} refer to a participant's use of spoken or written language in an activity, typically treated as externalizations of beliefs, goals, and plans; \textit{artifacts} refer to an object produced by a participant, such as a computer program. 
The think-aloud method can be used to collect accounts while carrying out an activity~\cite{tenenberg2019qualitative}. It can be used to gain insight into the cognitive processes of human problem-solving while performing a task, and can capture participants' strategies and intentions \cite{van1994think}. In this study, we use the think-aloud method to gain insight into students' reasonings when refactoring code. 



\begin{table*}[!htb]
\small
\sffamily
\caption{Code quality issues present in the refactoring exercises}
\label{tab:code-quality-issues}
\begin{tabular}{lll}
\hline
\rowcolor{light-blue}
\textbf{Code quality issue} & \textbf{In Exercises} & \textbf{Description} \\
\hline

Complex even number check & 5 & A suboptimal expression is used to check for even numbers \\ 

\rowcolor{very-light-blue}
Complex nested if statements & 4 & Two nested if statements with an else part are unnecessarily used \\ 

Duplicated statement & 1 & The same statement is written in both if and else parts \\ 

\rowcolor{very-light-blue}
God method & 4, 5 & A method has too many responsibilities \\ 

Long arithmetic expression & 1, 3 & An arithmetic expression is unnecessarily long \\ 

\rowcolor{very-light-blue}
Redundant boolean expression & 4, 5 & An unnecessary '== true' is used to check a boolean expression \\ 

Redundant if-else & 4 & An if-else statement returns a boolean literal directly \\ 

\rowcolor{very-light-blue}
Suboptimal loop choice & 2, 3, 4, 5 & A loop is used in place of another more suitable loop \\ 

Unnecessary else & 1, 3 & An unnecessary else part is used \\ 

\rowcolor{very-light-blue}
Unnecessary if & 2, 3 & An unnecessary if statement is used \\ 

\hline
\end{tabular}
\end{table*}

\subsection{Refactoring Exercises}

We designed five C\# programming exercises consisting of code that is functionally correct, but contains a number of code quality issues. The students' task was to refactor code to address these issues. They used DotNetFiddle\footnote{https://dotnetfiddle.net}, a free online code editor that contains only basic code features and does not provide suggestions for code refactoring.

Our exercises contain code quality issues reported in other studies, such as the redundant if-else \cite{da2023test, effenberger2022code}, duplicated statements \cite{da2023test, effenberger2022code}, unnecessary else \cite{da2023test}, and suboptimal loop choices \cite{effenberger2022code, tigina2023analyzing}. 
In total, we included ten different code quality issues. For each exercise, we provided a method containing between two and five issues -- see Table \ref{tab:code-quality-issues}. Some issues were present in multiple exercises. We included a description and test cases for each exercise. We also provided participants with a number of guidelines: do not change the header of a method; creating new methods is allowed; using built-in library methods such as \texttt{Sum()} or \texttt{Contains()} is not allowed; and do not address formatting. 

As an example, Exercise 3 contains the \texttt{GetDotPosition()} method. It takes a string \texttt{name} as argument and returns an integer, which represents the position of the first occurrence of the dot symbol "." in \texttt{name}. The method returns -1 in case \texttt{name} does not contain a dot. Empty strings do not need to be taken into account. Possible refactorings include replacing the while loop by a for loop, removing unnecessary conditions, and returning \texttt{pos} in the loop. Listing 1 shows an improved version. The other exercises can be found online.\footnote{https://sites.google.com/view/refactoring-exercises} Some exercises were inspired by examples in the literature. Exercise 2 was inspired by the \textit{CalculateScore} exercise from \citeauthor{keuning2021tutoring} \cite{keuning2021tutoring}, and Exercise 4 was inspired by a student solution for the \textit{SumValues} exercise from \citeauthor{luxton2013differences} \cite{luxton2013differences}.

\vspace{0.2cm}
{\centering
{\sffamily \small
Listing 1. Left: Refactoring Exercise 3. Right: An improved version of Exercise 3}

\begin{minipage}[t]{5.3cm}
\begin{lstlisting}[basicstyle={\footnotesize\ttfamily\linespread{1.0}\selectfont}, rulecolor=\color{black}, label={ref-ex-3}, frame=single]
int GetDotPosition(string name) {
    int pos = 0;
    while (pos < name.Length) {
        if (name[pos] == '.') {
            break;
        } else if (pos == name.Length-1) {
            pos = -1;
            break;
        } else {
            pos = pos + 1;
        }
    }
    return pos;
}
\end{lstlisting}
\end{minipage}
\hspace{4.0mm}
\begin{minipage}[t]{6.2cm}
\begin{lstlisting}[basicstyle={\footnotesize\ttfamily\linespread{1.0}\selectfont}, rulecolor=\color{teal}, label={improved-ref-ex-3}, frame=single]
int GetDotPosition(string name) {
    for (int pos = 0; pos < name.Length; pos++) {
        if (name[pos] == '.') {
            return pos;
        }
    }
    return -1;
}
\end{lstlisting}
\end{minipage}
\par}

\subsection{Participants}

We conducted a think-aloud study at Utrecht University, in the Netherlands. In this experiment, students worked on exercises in which they had to identify code quality issues, and apply refactorings to improve code. The university ethics board approved our study. We first organized a pilot study and tested it with three colleagues. The pilot study helped us to test and refine the settings for the actual experiment.

We recruited students by announcing the experiment in a lecture in the \textit{Modelling and System Development} (MSD) course. This is a mandatory second-year Computer Science course that covers, among other topics, software development methods, requirement analysis, object-oriented design, design patterns, unit testing, code quality and code refactoring. Our experiment was conducted before the lecture on code quality and code refactoring, to prevent any influence the lecture might have on students' behavior. A total of 12 students agreed to participate and signed an informed consent form. These students had already taken at least two programming courses: an introductory programming course, and a group project course. Aspects of code quality have been addressed in these courses, such as naming variables and commenting code.
C\# is the programming language used in both MSD and the prior programming courses. We summarize students' programming experience in Table \ref{tab:participants}.


\begin{table}
\small
\sffamily
\caption{Participants in the think-aloud study}
\begin{minipage}{0.48\linewidth}
\label{tab:participants}
\begin{tabular}{lclc}
\hline
\rowcolor{light-blue}
\textbf{Student} & \textbf{\makecell[c]{Year of \\ study}} & \textbf{Programming exp.} & \textbf{\makecell[c]{Years \\ of exp.}} \\
\hline
Student 01 & 3rd & Personal projects & 5 \\
\rowcolor{very-light-blue}
Student 02 & 2nd & University courses & 1 \\
Student 03 & 5th & Personal projects & 5 \\
\rowcolor{very-light-blue}
Student 04 & 4th & Double major CS \& Math & 1 \\
Student 05 & 4th & Majored in AI & 3 \\
\rowcolor{very-light-blue}
Student 06 & 2nd & Data analyst & 1 \\
\hline
\end{tabular}
\end{minipage}
\hspace{0.2cm}
\begin{minipage}{0.48\linewidth}
\begin{tabular}{lclc}
\hline
\rowcolor{light-blue}
\textbf{Student} & \textbf{\makecell[c]{Year of \\ study}} & \textbf{Programming exp.} & \textbf{\makecell[c]{Years \\ of exp.}} \\
Student 07 & 2nd & Personal projects & 4 \\
\rowcolor{very-light-blue}
Student 08 & 2nd & Programming tutor & 5 \\
Student 09 & 2nd & Web developer & 6 \\
\rowcolor{very-light-blue}
Student 10 & 4th & Physics student, CS minor & 1 \\
Student 11 & 2nd & University courses & 1 \\
\rowcolor{very-light-blue}
Student 12 & 2nd & Game developer & 5 \\
\hline
\end{tabular}
\end{minipage}
\end{table}

\subsection{Think-Aloud Procedure}
\label{think-aloud-procedure}

The first author carried out individual in-person think-aloud sessions with the participants. In each session, the participant used their own computer and was invited to join an online call on Microsoft Teams. Each session lasted approximately one hour, and had the following structure:

\textit{Warm-up} (15 minutes): This part was not recorded. The researcher explained the activity goal -- to remove code quality issues through code refactoring -- to the student, without further explaining code refactoring. The student received preliminary instructions and practiced the think-aloud method with a sample programming exercise.

\textit{Think-aloud session} (30 minutes): The recording of the session started. The student shared their screen and worked on five refactoring exercises while expressing their ideas. Since the experiment asked participants to think aloud while programming, both the student's voice and screen actions were recorded. The researcher observed the student's behavior and made note of unclear verbalizations and actions.

\textit{Wrap-up} (15 minutes): This part was not recorded. The student was given the opportunity to ask questions and offer feedback on the session. The student and researcher discussed code quality issues and potential refactoring steps for the exercises. If necessary, the researcher asked for clarification of particular verbalizations and actions.

\subsection{Data Transcription}

The first author used the recordings to transcribe student verbalizations and log their code snapshots into a spreadsheet. For privacy reasons, only this researcher had access to the recordings. To facilitate the analysis of our dataset, we used ProgSnap2 \cite{price2020progsnap2}, a standardized format for representing programming snapshot data. Besides the required fields from ProgSnap2, we included other fields to represent and analyze our data. Following the format, these new fields start with \textit{X-}. We added a field \textit{X-Verbalization} to represent student and interviewer fragments of verbalization during the think-aloud session. Whenever a student paused and resumed their verbalization or spoke about a different section of code, a new row was added in the spreadsheet.

\subsection{Data Analysis}



\subsubsection{Student actions in code refactoring}
\label{student-actions}

We use qualitative thematic analysis to code our dataset. In the first step, we identify categories of student actions while working on the exercises. This categorization facilitates the thematic analysis. The first two authors analyzed four full sequences of snapshots from different students working on an exercise, and classified their actions. The first author later used this categorization with two other full sequences of snapshots from different students, and no new categories emerged.

We identify five categories of student actions: \textit{editing} means that a student is changing code; \textit{testing} means that a student is compiling code, checking test cases or program output; \textit{reading} means that a student is reading the exercise prompt or a fragment of code; \textit{reflecting} is a broad term to represent student reasoning. It includes situations in which a student thinks aloud about code semantics, chooses to perform a refactoring step, identifies an issue present in the exercise, or tries to understand code; \textit{other speech} means that a student is talking about something else, such as finishing one exercise and moving on to the next. We added a field \textit{X-RefAction} to the dataset to represent the coding of a student action. If a student performs multiple actions simultaneously -- e.g.~\textit{editing} code and \textit{reflecting} -- we include both actions in the same cell. We code these actions separately.

\subsubsection{RQ1. Students' reasoning in code refactoring}
\label{analysis-RQ1}



The Handbook of Computing Education Research \cite{tenenberg2019qualitative} presents \textit{inductive categorization} as an appropriate qualitative method for studies focusing on understanding what categories subdivide an object of study. This method is often used in studies in which no categorization exists for such an object.
As far as we are aware, prior studies in Computing Education have not investigated student reasoning when working on code refactoring tasks. Therefore, we perform an inductive analysis of this topic.
For this, we analyzed every occurrence of a \textit{reflecting} action in the dataset. We adopted a grounded-theory based approach following the coding phases from \citeauthor{strauss1990basics} \cite{strauss1990basics}, in which all three authors participated in the coding process.

In the \textit{open coding} phase, the first author randomly selected four sequences of snapshots, coded all rows in which a reflection occurred, and discussed this preliminary coding with the other researchers. From this discussion, we developed an initial coding of student reasoning.

In the \textit{axial coding} phase, the first author randomly selected four different sequences of snapshots. These  were separately coded by the first two authors and then discussed by all authors. Subsequently, we repeated this process with four other sequences from different students and exercises. We refined the coding in multiple rounds of discussion.

In \textit{selective coding}, we merged similar categories from the previous phase to obtain a final coding. We fit all existing categories and no new categories emerged when coding the snapshots since the seventh student, which indicated theoretical saturation \cite{hennink2022sample, strauss1998basics}. The labels from the coding were included in our dataset in the \textit{X-Reasoning} field.



\subsubsection{RQ2. Relation between addressing code quality issues and reasoning while refactoring}

To answer RQ2, we first analyzed which code quality issues were addressed. Whenever a student mentioned or edited code to address an issue, we registered this in two fields: \textit{X-CodeQualityIssue} indicates which issue is addressed; \textit{X-CodeIssueStatus} represents the status of the issue in each row: \textit{Identified}: a student mentions the presence of an issue, but does not refactor code to address it; \textit{Fixed}: a student removes an issue through refactoring; \textit{Improved}: a student takes a helpful refactoring step towards removing an issue, but the issue is still present; \textit{Attempted}: a student attempts to remove an issue, but the change has no effect on code quality; \textit{Worsened}: a student takes an unhelpful step while working on an issue, decreasing the code quality. In some cases, addressing an issue required multiple refactoring steps (\textit{editing} actions). For example, replacing a while loop by a for loop in Exercise 3 requires deleting the while, updating the if statements, and writing the for loop. In our spreadsheet, we labeled the final step of this sequence as the one that fixed an issue. Next, we compared the issues addressed by students with their reasonings (X-RefAction field). We identified which reasons were given to address each code quality issue, and whether those issues were fixed.





\section{Results}
\label{Results}



\subsection{RQ1. Students' reasoning in code refactoring}
\label{results-rq1}

\begin{table*}[]
\small
\centering
\sffamily
\caption{Categories of student reasoning in code refactoring}
\label{tab:reasoning}
\begin{tabular}{lllr}

\hline
\textbf{Reasoning} & \textbf{Description} & \textbf{Example of student reasoning} & \textbf{Occ.} \\
\hline
\rowcolor{light-blue}
\multicolumn{4}{l}{\textit{\textbf{Presence of code quality issues}}} \\

Unnecessary Code & \makecell[l]{Student reasons about a fragment of code \\ that is not necessary and can be removed.} & \makecell[l]{\textit{"I'd take it outside of [if] statements, then I could} \\ \textit{remove the entire else because nothing happens."}} & 61 \\
\rowcolor{very-light-blue}
Complex Code Structure & \Gape[1pt][2pt]{\makecell[l]{Student reasons about a code structure \\ that is too complex and can be simplified.}} & \Gape[0pt][2pt]{\makecell[l]{\textit{"This while loop part is too complicated, and} \\ \textit{it behaves like a for loop. I’ll change it."}}} & 18 \\
Long Method & \makecell[l]{Student reasons about a method that has \\ too many responsibilities.} & \makecell[l]{\textit{"Everything is inside one method! I’ll create a} \\ \textit{second method to check if it [a number] is prime."}} & 7 \\

\rowcolor{light-blue}
\multicolumn{4}{l}{\textit{\textbf{Improvement of code quality attributes}}} \\
\rowcolor{very-light-blue}
Improve Readability & \Gape[1pt][2pt]{\makecell[l]{Student reasons about a refactoring step \\ to improve code readability.}} & \Gape[0pt][2pt]{\makecell[l]{\textit{"I'll remove the while loop and write a for [loop].} \\ \textit{This will make code easier to read."}}} & 7 \\

Improve Performance &  \Gape[1pt][2pt]{\makecell[l]{Student reasons about a refactoring step \\ to improve code performance.}} & \Gape[1pt][2pt]{\makecell[l]{\textit{"This method can be faster if the loop starts from} \\ \textit{the second iteration."}}} & 5 \\

\rowcolor{very-light-blue}
Improve Maintainability & \Gape[1pt][2pt]{\makecell[l]{Student reasons about a refactoring step \\ to improve code maintainability.}} & \makecell[l]{\textit{"I can just create a salary [variable]. It makes code} \\ \textit{more (...) robust to changes."}} & 3 \\

\rowcolor{light-blue}
\multicolumn{4}{l}{\textit{\textbf{Code semantics}}} \\
Code Comprehension & \Gape[1pt][2pt]{\makecell[l]{Student reasons about the purpose of the \\ code.}} & \makecell[l]{\textit{"It [the method] starts with the 24 hours of a day.} \\ \textit{So if it is July or August... Just return 24 hours."}} & 11 \\
\rowcolor{very-light-blue}
Semantic Preservation & \Gape[1pt][2pt]{\makecell[l]{Student reasons about code correctness \\ after a refactoring.}} & \Gape[0pt][2pt]{\makecell[l]{"\textit{It works, but it depends on the stuff in the array} \\ \textit{(the parameter) to be larger than this value."}}} & 6 \\
\hline
\end{tabular}
\end{table*}

We identified eight categories of student reasonings in code refactoring. 
Three of these reason about the presence of code quality \textit{issues} (see definition in \ref{Definitions}), three others about the improvement of code quality \textit{attributes}, and two about code semantics. A summary of the categories of students' reasonings is shown in Table \ref{tab:reasoning}, including a short description, an example and the number of occurrences of each reasoning. In the examples below, a black frame represents the original exercise code or an intermediate state. A green frame represents a correct state. A red frame represents a refactoring that decreased code quality or changed its semantics.



\vspace{3mm}

\textit{\textbf{Presence of code quality issues.}} A student found a quality issue in code, and thinks about possible ways to remove it. We identified three categories for this type of reasoning.

\vspace{3mm}

$\bullet$ \textit{Unnecessary Code.} A fragment of code is not necessary and can be removed. Other terms students often use are \textit{not needed}, \textit{irrelevant}, \textit{redundant}, \textit{superfluous}. For example, in Exercise 1 Student 11 notices:

\vspace{5mm}

{\centering
\fcolorbox{white}{very-light-blue}{\begin{minipage}{12cm}
\textit{"It says 'freeHours - studyingHours', but it's zero anyway! Why does that [subtraction] matter?"}
\end{minipage}} \\[2mm]
\begin{minipage}{4.6cm}
\begin{lstlisting}[basicstyle={\footnotesize\ttfamily\linespread{1.0}\selectfont}, frame=single, rulecolor=\color{black}]
int freeHours = 24;
int studyingHours = 0;
if (month == 7 || month == 8) {
    return freeHours - studyingHours;
} else {
    ...
}
\end{lstlisting}
\end{minipage}
\hspace{0.2cm}
\textrightarrow
\hspace{0.3cm}
\begin{minipage}{4.0cm}
\begin{lstlisting}[basicstyle={\footnotesize\ttfamily\linespread{1.0}\selectfont}, frame=single, rulecolor=\color{teal}]
int freeHours = 24;
int studyingHours = 0;
if (month == 7 || month == 8) {
    return freeHours;
} else {
    ...
}
\end{lstlisting}
\end{minipage} \\[8mm]
\par}


Another example of \textit{unnecessary code} reasoning in Exercise 1 is given by Student 01, who correctly extracts the duplicated return statement from the if and else parts. The student moves the calculation of \texttt{studyingHours} to the if part, and reasons about removing the empty else. Even though the reasoning is adequate, the student takes an incorrect step when updating the condition. The == operators are correctly replaced by != operators, but the or-operator is not replaced by an and-operator. This incorrect step is an instance of a refactoring misconception, \textit{BadIfElseSimplification}, identified in our prior study \cite{oliveira2023student}.

\vspace{3mm}

{\centering
\fcolorbox{white}{very-light-blue}{\begin{minipage}{14cm}
\textit{"I would take it outside of the [conditional] statements, then I could remove the entire else because nothing happens."}
\end{minipage}} \\[2mm]
\begin{minipage}{4.5cm}
\begin{lstlisting}[basicstyle={\footnotesize\ttfamily\linespread{1.2}\selectfont}, frame=single, rulecolor=\color{black}]
int freeHours = 24;
int studyingHours = 0;
if (month == 7 || month == 8) {
  return freeHours - studyingHours;
} else {
  studyingHours += (courses * 3);
  return freeHours - studyingHours;
}
\end{lstlisting}
\end{minipage}
\hspace{0.1cm}
\textrightarrow
\hspace{0.2cm}
\begin{minipage}{4.3cm}
\sethlcolor{light-green}
\begin{lstlisting}[basicstyle={\footnotesize\ttfamily\linespread{1.2}\selectfont}]
int freeHours = 24;
int studyingHours = 0;
if (month == 7 || month == 8) {
} else {
  studyingHours += (courses * 3);
}
return freeHours - studyingHours;
\end{lstlisting}
\end{minipage}
\hspace{0.1cm}
\textrightarrow
\hspace{0.2cm}
\begin{minipage}{4.3cm}
\sethlcolor{light-green}
\begin{lstlisting}[basicstyle={\footnotesize\ttfamily\linespread{1.2}\selectfont}, frame=single, rulecolor=\color{red}]
int freeHours = 24;
int studyingHours = 0;
if (month != 7 || month != 8) {
  studyingHours += (courses * 3);
}
return freeHours - studyingHours;
\end{lstlisting}
\end{minipage} \\[1cm]
\par}

$\bullet$ \textit{Complex Code Structure.} A code structure (e.g.~a loop or a condition) is too complex, and it can be simplified. A few common terms used by students are \textit{complex} and \textit{complicated}. In the example below from Exercise 3, a while loop is used to iterate through a string. Student 06 notices that using a for loop would be a less complicated choice.

\vspace{3mm}

{\centering
\fcolorbox{white}{very-light-blue}{\begin{minipage}{10.5cm}
\textit{"This while loop part is too complicated, and it behaves like a for loop. I'll change it."}
\end{minipage}} \\[2mm]
\begin{minipage}{3.6cm}
\begin{lstlisting}[basicstyle={\footnotesize\ttfamily\linespread{1.0}\selectfont}, frame=single, rulecolor=\color{black}]
int pos = 0;
while (pos < name.Length) {
    if (name[pos] == '.') {
        return pos;
    }
    pos = pos + 1;
}
return -1;
\end{lstlisting}
\end{minipage}
\hspace{0.2cm}
\textrightarrow
\hspace{0.3cm}
\begin{minipage}{5.8cm}
\begin{lstlisting}[basicstyle={\footnotesize\ttfamily\linespread{1.0}\selectfont}, frame=single, rulecolor=\color{teal}]
for (int pos = 0; pos < name.Length; pos++) {
    if (name[pos] == '.') {
        return pos;
    }
}
return -1;
\end{lstlisting}
\end{minipage} \\[8mm]
\par}

Another example of this reasoning is given by Student 07, who reasons in the same exercise about the complex logic for the case where a dot is not found in a string. However, the student takes an inadequate refactoring step, in which they remove the else if part and write an unnecessary condition out of the loop. The students says:

\vspace{5mm}

{\centering
\fcolorbox{white}{very-light-blue}{\begin{minipage}{12cm}
\textit{"This check if you are at the end [of the string] can be simplified. You can write it out of the while."}
\end{minipage}} \\[2mm]
\begin{minipage}{4.8cm}
\begin{lstlisting}[basicstyle={\footnotesize\ttfamily\linespread{1.0}\selectfont}, frame=single, rulecolor=\color{black}]
while (pos < name.Length) {
    ...
    } else if (pos == name.Length-1) {
        pos = -1;
        break;
    }
    ...
}
return pos;
\end{lstlisting}
\end{minipage}
\hspace{0.1cm}
\textrightarrow
\hspace{0.2cm}
\begin{minipage}{7.0cm}
\begin{lstlisting}[basicstyle={\footnotesize\ttfamily\linespread{1.0}\selectfont}, frame=single, rulecolor=\color{red}]
int pos = 0;
while (pos < name.Length) {
    ...
}
if (pos == name.Length && name[name.Length-1] != '.') {
    pos = -1;
}
return pos;
\end{lstlisting}
\end{minipage} \\[1cm]
\par}

$\bullet$ \textit{Long Method.} A method has too many responsibilities. For example, Student 06 realizes in Exercise 5 that the method has multiple responsibilities, and decides to break it into two smaller methods: one to detect whether a number is prime, and another to count the prime numbers in the array.

\vspace{3mm}

{\centering
\fcolorbox{white}{very-light-blue}{\begin{minipage}{12cm}
\textit{"Everything is inside one method! I'll create a second method to check if it [a number] is prime."}
\end{minipage}} \\[2mm]
\begin{minipage}{5.9cm}
\begin{lstlisting}[basicstyle={\footnotesize\ttfamily\linespread{1.0}\selectfont}, frame=single, rulecolor=\color{black}]
static int CountPrimeNumbers(int[] numbers) {
    int count = 0;
    bool IsPrime;
    for (int i = 0; i < numbers.Length; i++) {
        IsPrime = true;    
        for (int j = 2; j ... ; j++) {
            ...
        }            
        ...
    }
    return count;
}
\end{lstlisting}
\end{minipage}
\hspace{0.2cm}
\textrightarrow
\hspace{0.3cm}
\begin{minipage}{5.9cm}
\begin{lstlisting}[basicstyle={\footnotesize\ttfamily\linespread{1.0}\selectfont}, frame=single, rulecolor=\color{teal}]
static int CountPrimeNumbers(int[] numbers) {
    int count = 0;
    for (int i = 0; i < numbers.Length; i++) {
        if (IsPrime(numbers[i]))
            count += 1;
    }
    return count;
}

static bool IsPrime(int n) {
    ...
}
\end{lstlisting}
\end{minipage} \\[7mm]
\par}

\textit{\textbf{Improvement of code quality attributes.}} 
We identified the following three categories.

\vspace{3mm}

$\bullet$ \textit{Improve Readability.} A few students suggest that code can be modified to improve readability. 
For instance, Student 05 realizes in Exercise 3 that the while loop is a less suitable loop choice -- the same example as given under \textit{Complex Code Structure}, but the reasoning presented in that case was different. Student 05 reflects on replacing the loop:

\vspace{3mm}

{\centering
\fcolorbox{white}{very-light-blue}{\begin{minipage}{10.5cm}
\textit{"I'll remove the while loop and write a for [loop]. This will make code easier to read."}
\end{minipage}} \\[7mm]
\par}

$\bullet$ \textit{Improve Performance.} One reason given by students to refactor code is to improve performance. Some common terms used by students are \textit{faster} and \textit{more efficient}. In the following example, Student 07 works on Exercise 2 and observes that the for loop starts the iteration from the first element of the array. Since the first element is already assigned to \texttt{largest}, this iteration can be skipped, so that the for loop starts the iteration from the second element.

\vspace{3mm}

{\centering
\fcolorbox{white}{very-light-blue}{\begin{minipage}{9cm}
\textit{"This method can be faster if the loop starts from the second iteration."}
\end{minipage}} \\[2mm]
\begin{minipage}{5.3cm}
\begin{lstlisting}[basicstyle={\footnotesize\ttfamily\linespread{1.0}\selectfont}, frame=single, rulecolor=\color{black}]
int largest = numbers[0];
for (int i = 0; i < numbers.Length; i++) {
    ...
}
return largest;
\end{lstlisting}
\end{minipage}
\hspace{0.2cm}
\textrightarrow
\hspace{0.3cm}
\begin{minipage}{5.3cm}
\sethlcolor{light-green}
\begin{lstlisting}[basicstyle={\footnotesize\ttfamily\linespread{1.0}\selectfont}, frame=single, rulecolor=\color{teal}]
int largest = numbers[0];
for (int i = 1; i < numbers.Length; i++) {
    ...
}
return largest;
\end{lstlisting}
\end{minipage} \\[1cm]
\par}

$\bullet$ \textit{Improve Maintainability.} Students mention that changing code would improve maintainability. In this category, we observed that in all cases \textit{readability} was also mentioned along with maintainability. For example, in Exercise 4, Student 09 thinks about creating a \texttt{salary} variable to temporarily store the value of a salary when iterating over the \texttt{salaries} array, and mentions that this step makes code more robust to changes -- which improves maintainability. 

\vspace{3mm}

{\centering
\fcolorbox{white}{very-light-blue}{\begin{minipage}{11cm}
\textit{"I can just create a salary [variable]. It makes code more readable and robust to changes."}
\end{minipage}} \\[2mm]
\begin{minipage}{5.5cm}
\begin{lstlisting}[basicstyle={\footnotesize\ttfamily\linespread{1.0}\selectfont}, frame=single, rulecolor=\color{black}]
float sum = 0.0F;
for (int i = 0; i < salaries.Length; i++) {
    if (salaries[i] <= 15000 || ...) {
        sum += salaries[i];               
    }
}
\end{lstlisting}
\end{minipage}
\hspace{0.2cm}
\textrightarrow
\hspace{0.3cm}
\begin{minipage}{5.5cm}
\sethlcolor{light-green}
\begin{lstlisting}[basicstyle={\footnotesize\ttfamily\linespread{1.0}\selectfont}, frame=single, rulecolor=\color{teal}]
float sum = 0.0F;
for (int i = 0; i < salaries.Length; i++) {
    float salary = salaries[i];
    if (salary <= 15000 || ...) {
        sum += salary;
    }
}
\end{lstlisting}
\end{minipage} \\[7mm]
\par}

\textit{\textbf{Code semantics.}} 
We identified the two categories below.

\vspace{3mm}

$\bullet$ \textit{Code Comprehension.} A student tries to understand what the code does. All students spent some time reasoning about the code's purpose. For example, Student 12 works on Exercise 1 and reasons about how the program works.

\vspace{3mm}

{\centering
\fcolorbox{white}{very-light-blue}{\begin{minipage}{12cm}
\textit{"It [the method] starts with the 24 hours of a day. So if it is July or August... Just return 24 hours."}
\end{minipage}} \\[7mm]
\par}

$\bullet$ \textit{Semantics Preservation.} A student thinks about whether or not the code would (still) be semantically correct, or reflects on test results. For example, in Exercise 2 Student 03 first initializes the \texttt{largest} variable with 0. As a consequence, some test cases that contain only negative numbers fail. The student eventually realizes the semantic issue, then changes the initial value of \texttt{largest} to -65000. The student incorrectly thinks that this number is close to the minimum int value. The student runs the test cases again, and reflects on the results:

\vspace{3mm}

{\centering
\fcolorbox{white}{very-light-blue}{\begin{minipage}{12cm}
\textit{"It works, but it depends on the stuff in the array [the parameter] to be larger than this value."}
\end{minipage}} \\[2mm]
\begin{minipage}{5.3cm}
\begin{lstlisting}[basicstyle={\footnotesize\ttfamily\linespread{1.0}\selectfont}, frame=single, rulecolor=\color{red}]
int largest = -65000;
for (int i = 0; i < numbers.Length; i++) {
    if (numbers[i] > largest) {
        largest = numbers[i];
    }
}    
return largest;
\end{lstlisting}
\end{minipage} \\[1cm]
\par} 

\textit{\textbf{Other verbalizations.}} In several cases, students refactor code without giving a reason. They either verbalize very little or literally describe the steps they take. This mostly occurs when students are refactoring conditional statements or loops.
For example, in Exercise 4, Student 08 correctly identifies that the block with two nested if statements and an else part can be simplified. However, without providing an explanation, the student merges both if statements with an and-operator and did not take the else part into account -- this incorrect refactoring step is an instance of the \textit{IfsMerged} refactoring misconception~\cite{oliveira2023student}. The student just describes their refactoring, without providing a reason. The student then notices the issue, reverts the refactoring, and moves on to another exercise.

\vspace{3mm}

{\centering
\fcolorbox{white}{very-light-blue}{\begin{minipage}{6.5cm}
\textit{"There is a nested if right there. This can just be one."}
\end{minipage}} \\[2mm]
\begin{minipage}{3.8cm}
\begin{lstlisting}[basicstyle={\footnotesize\ttfamily\linespread{1.0}\selectfont}, frame=single, rulecolor=\color{black}]
if (skipLowSalary) {
    if (salaries[i] > 15000) {
        sum += salaries[i];
    }
} else {
    sum += salaries[i];
}
\end{lstlisting}
\end{minipage}
\hspace{0.2cm}
\textrightarrow
\hspace{0.3cm}
\begin{minipage}{5.5cm}
\sethlcolor{light-red}
\begin{lstlisting}[basicstyle={\footnotesize\ttfamily\linespread{1.0}\selectfont}, frame=single, rulecolor=\color{red}]
if (skipLowSalary && salaries[i] > 15000) {
    sum += salaries[i];
} else {
    sum += salaries[i];
}
\end{lstlisting}
\end{minipage} \\[7mm]
\par}



\subsection{RQ2. Relation between addressing code quality issues and reasoning while refactoring}

{\centering
\begin{table*}[ht]
\small
\sffamily
\caption{Students addressing code quality issues - Final states. \textit{Iden.:} an issue identified, but not addressed; \textit{Fixed:} the removal of an issue through refactoring; \textit{Impr.:} a helpful step towards the removal of an issue; \textit{Tried:} an attempt to remove an issue which had no effect in code; \textit{Wors.:} a step that decreased code quality and did not remove an issue; \textit{Not Id.:} an issue not identified.}
\label{tab:code-quality-issues-final-status}
\begin{tabular}{llr|rrrrrr}
\hline
\rowcolor{light-blue}
\textbf{Code quality issue} & \textbf{Reasonings used} & \textbf{Occ.} & \textbf{Iden.} & \textbf{Fixed} & \textbf{Impr.} & \textbf{Tried} & \textbf{Wors.} & \textbf{Not Id.} \\
\hline
Complex even number check & \makecell[l]{Improve performance (2) \\ Complex code structure (1)} & 9 & 0 & 3 & 0 & 0 & 0 & 6 \\
\rowcolor{very-light-blue}
Complex nested if statements & \makecell[l]{Unnecessary code (4) \\ Complex code structure (2) \\ Improve readability (1)} & 11 & 1 & 2 & 3 & 2 & 1 & 2 \\
Duplicated statement & \makecell[l]{Unnecessary code (8) \\ Improve performance (2)} & 12 & 1 & 8 & 0 & 1 & 1 & 1 \\
\rowcolor{very-light-blue}
God method & \makecell[l]{Long method (7) \\ Improve readability (1) \\ Improve maintainability (1)} & 20 & 0 & 7 & 1 & 0 & 0 & 12 \\
Long arithmetic expression & \makecell[l]{Complex code structure (7) \\ Improve readability (3)} & 24 & 0 & 20 & 0 & 0 & 0 & 4 \\
\rowcolor{very-light-blue}
Redundant boolean expression & Unnecessary code (12) & 20 & 0 & 19 & 0 & 0 & 0 & 1 \\
Redundant if-else & \makecell[l]{Complex code structure (5) \\ Unnecessary code (2)} & 11 & 0 & 10 & 1 & 0 & 0 & 0 \\
\rowcolor{very-light-blue}
Suboptimal loop choice & \makecell[l]{Complex code structure (3) \\ Improve performance (2) \\ Improve readability (2) \\ Improve maintainability (1)} & 42 & 0 & 10 & 0 & 0 & 0 & 32 \\
Unnecessary else & Unnecessary code (6) & 12 & 0 & 8 & 0 & 1 & 1 & 2 \\
\rowcolor{very-light-blue}
Unnecessary if & Unnecessary code (29) & 42 & 0 & 34 & 0 & 3 & 3 & 2 \\
\textbf{Total} & - & \makecell[r]{203 \\ (100\%)} & \makecell[r]{2 \\ (1\%)} & \makecell[r]{121 \\ (60\%)} & \makecell[r]{5 \\ (2\%)} & \makecell[r]{7 \\ (3\%)} & \makecell[r]{6 \\ (3\%)} & \makecell[r]{62 \\ (31\%)} \\
\hline
\end{tabular}
\end{table*}
\par}

To answer RQ2, 
we inspected the sequences of program snapshots from 12 students, who worked on five exercises containing a total of 10 different code quality issues. Each sequence represents an exercise attempted by a student. 
We analyzed 54 sequences to find out what code quality issues were addressed. Not every student attempted all exercises. A few issues were present in multiple exercises, such as the \textit{suboptimal loop choice}. We summarize our findings in Table \ref{tab:code-quality-issues-final-status}.

Students could have encountered 203 issues in the exercises they worked on. We found that they worked on 141 of those, and did not identify 62 issues -- see the \textit{Not Id.} column in Table \ref{tab:code-quality-issues-final-status}. An issue was fixed in 60\% of the 203 cases (\textit{Fixed} column). 2\% of the steps were helpful for removing of an issue, but did not remove the issue completely (\textit{Impr.} column). In 3\% of the cases, students attempted to remove an issue (\textit{Tried} column) but their steps had no effect in terms of quality or semantic. 3\% of the steps decreased code quality (\textit{Wors.} column). In two occasions, students mentioned a code quality issue (\textit{Iden.} column), but did not take a refactoring step to address it. Other refactoring steps, such as removing brackets, were not related to the issues present in the exercises.



Two code quality issues were frequently overlooked: the \textit{god method} issue was not identified in 12 occasions, and the \textit{suboptimal loop choice} was not identified in 32 occasions. Six students fixed the \textit{god method} issue in the cases where they did identify it, which often occurred in Exercise 5. Two students mentioned that the method in Exercise 4 had too many responsibilities, and split it into two smaller methods: one method to calculate the salaries, and another one to check if the salary limit has been reached. The \textit{suboptimal loop choice} issue was mostly overlooked in the exercises where a for loop could be replaced by a foreach loop. It was fixed only in two out of 30 cases. In the discussions in the wrap-up phase, many students said that they would not consider replacing a for loop by a foreach loop, but would use a foreach loop when writing code from scratch. The other instance of \textit{suboptimal loop choice} concerned replacing a while loop by a for loop. This issue was fixed in eight out of 12 cases.

The \textit{Reasonings used} column in Table \ref{tab:code-quality-issues-final-status} presents the reasons given by students when addressing a code quality issue. In 43 cases, students addressed an issue without giving a reason.
The most commonly removed code quality issues are related to the \textit{Unnecessary Code} reasoning. For example, the \textit{redundant boolean expression} was fixed in all 19 occurrences where students identified it, and 12 students explained that the second part of the expression was unneeded. Nevertheless, the \textit{Unnecessary Code} reasoning is also often used to motivate incorrect refactoring steps. For example, in an attempt to remove the \textit{duplicated statement} issue in Exercise 1, a student incorrectly verbalized that the whole if-else block was not necessary, and removed it. This incorrect refactoring step is an instance of the \textit{BadIfElseSimplification} misconception. Moreover, the \textit{Unnecessary Code} reasoning was related to incorrect refactoring attempts to remove the \textit{complex nested if statements} in Exercise 4. Three students justified that the inner if statement was not necessary and removed it, without taking the else part into account. This incorrect step is an instance of the \textit{MergedIfs} refactoring misconception.


We observed that students perform the same refactorings, but express different reasons. This occurs in cases involving the improvement of code readability. E.g., in a refactoring where a while loop is replaced by a for loop, a student mentions the presence of a while loop as an issue to be fixed, whereas another student motivates this refactoring as a way to improve readability. 
We also observed that pointing at a code quality issue was more common (86 occurrences) than reasoning about the improvement of a code quality attribute (15 occurrences).
In 13 out of 15 cases, students took correct refactoring steps, i.e.~a step that resulted in an \textit{improved} or \textit{fixed} issue, when reasoning about the improvement of a quality attribute.

\section{Discussion}
\label{Discussion}


This section discusses the most relevant findings for our research questions.

\vspace{0.05cm}

\textbf{Our collection of reasonings relates to subsets from other classifications.} We found eight categories of student reasonings when refactoring code. The reasonings from the \textit{Presence of code quality issues} category relate to a subset of defect types identified in \citeauthor{vrechtavckova2024catalog}'s \cite{vrechtavckova2024catalog} study. They present a catalog of code quality issues (called \textit{defects}), categorized into eight types. The \textit{Unnecessary code} reasoning relates to the \textit{unused code} and \textit{duplicate code} defect types; the \textit{Complex code structure} reasoning relates to \textit{simplifiable}; and the \textit{Long method} reasoning relates to \textit{poor design}. Two of the reasonings from the \textit{Improvement of code quality attributes} category relate to properties from the ISO/IEC 25010 quality model \cite{ISO25010}. The \textit{Improve performance} and \textit{Improve maintainability} reasonings relate to the \textit{performance efficiency} and \textit{maintainability} properties. Not all defect types from \citeauthor{vrechtavckova2024catalog}'s catalogue appeared in our students' reasonings, which might be attributed to the fact that we told our students not to focus on formatting issues, or students just used other terminology. In addition, many of the quality attributes were not relevant to our exercises. Because we categorized student's reasonings, and not code issues or attributes, we could not use these existing classifications as a basis for our classification.

\vspace{0.1cm}

\textbf{Experienced students reason about quality \textit{attributes} more often.} Six students explicitly mentioned an improvement in readability, maintainability or performance as a reason to refactor code, rather than pointing at a problem they see in the code. 
In \citeauthor{izu2023exploring}'s \cite{izu2023exploring} study, students ranked a set of code solutions in terms of quality and gave explanations for their choices. Performance and structure were the most common quality attributes mentioned by students, whereas readability was only the fifth most cited attribute. In our study, reasoning about quality attributes occurred more frequently among students with more programming experience. Moreover, we observed that students gave different reasons to perform the same refactorings. This occurred in cases where a student motivated their refactoring with an improvement of code readability, whereas another student pointed out an issue, such as unnecessary or complex code.
We think that students who reason about quality attributes are more used to work on group projects, in which code quality is more important, whereas students who reason about quality issues are less experienced and tend to focus on the issues they see in code.
Regardless of the student experience level, reasoning about quality attributes occurred along with taking correct refactoring steps in 13 out 15 cases. Regarding the reasoning about code semantics, we observed that three students were concerned about preserving semantically correct code after each refactoring step. This is similar to the finding from \citeauthor{da2023test}'s \cite{da2023test} study.

\vspace{0.1cm}

\textbf{Students have different perceptions and preferences for code readability.} We noticed that the term \textit{readability} is used to describe different aspects of code quality, such as layout and structure. For example, one student replaced a while loop by a for loop in Exercise 3 and mentioned an improvement in readability as a reason for the refactoring; one student removed brackets whenever possible in every exercise and also mentioned readability improvement as motivation for taking this step. Students' preferring a for loop over the while loop was found in \citeauthor{wiese2019linking}'s \cite{wiese2019linking} study, in which students were exposed to two equivalent programs using these loop choices, and 84\% of them considered the for loop version more readable. Concerning the \textit{redundant if-else} issue, 95\% of students refactored the if-else statement to return a boolean expression directly. Three students mentioned the presence of \textit{complex code structure} as a reason to refactor the if-else statement. Our finding differs from \citeauthor{wiese2019replicating}'s \cite{wiese2019replicating} study, which found that 66\% of students preferred an if statement to return true or false rather than directly returning a boolean expression. Our result might have been affected by the example exercise used in the warm-up phase, which contained the same quality issue.

\vspace{0.1cm}

\textbf{Students are concerned about code performance, and this may affect other quality attributes.} We observed that performance is a common concern among students, and they may refactor code to obtain marginal performance improvements. We noticed one example in Exercise 5. Two students who worked on this exercise dedicated some time reasoning about the most efficient way to calculate prime numbers. In Exercise 2, two students changed the loop to start from the second iteration, since the variable largest (outside the loop) was already assigned the first element of the array, and the first iteration could be skipped. One example of refactoring to improve performance may have affected other quality attributes: in Exercise 1, two students removed the freeHours variable and replaced its occurrences with a literal value (24) -- an instance of the \textit{magic number} code smell \cite{beck2018refactoring}. This refactoring is undesirable, because it decreases code readability and maintainability.

\vspace{0.1cm}

\textbf{The \textit{god method} and \textit{suboptimal loop choice} issues were often overlooked.} We observed that students removed issues in 60\% of the cases. They were particularly successful with removing issues related to redundant code: in 95\% of cases in the \textit{redundant boolean expression}, and 91\% of cases in the \textit{redundant if-else} issue. This finding supports the results from \citeauthor{tigina2023analyzing}'s \cite{tigina2023analyzing} and \citeauthor{keuning2020student}'s \cite{keuning2020student} studies, which also found that most students fix boolean expression issues. However, our results show that the \textit{suboptimal loop choice} issue was overlooked in 76\% of cases. This finding differs from \citeauthor{tigina2023analyzing}'s \cite{tigina2023analyzing} and \citeauthor{keuning2020student}'s \cite{keuning2020student} studies, in which students replaced a for loop by a foreach loop in most cases. We think that this may be related to the hints offered by tools used in their experiments. In \citeauthor{keuning2020student}'s study, for example, 61\% of students viewed a hint on this issue. Two exercises contained the \textit{god method} quality issue. Although this issue is commonly associated with object oriented-level programs, it can also be found in smaller programs at the method-level \cite{effenberger2022code}. We observed that 50\% of students fixed or improved the \textit{god method} in Exercise 5, but only 16\% of them fixed the issue in Exercise 4. This difference does not seem to be related to the number of lines of code, since both exercises contain a single method with 17 lines. Our findings relate to \citeauthor{nurollahian2023use}'s \cite{nurollahian2023use} study, which found the use of both \textit{correct patterns} and \textit{anti-patterns} in the code of the same student. In our study, based on observations of students' reactions when working on Exercise 5 and the conversations in the wrap-up phase, we think that their decision to split the methods only in that exercise is related to the fact that they have worked on similar exercises involving prime numbers before. 

\vspace{0.1cm}

\textbf{Students' difficulties with conditional statements are instances of refactoring misconceptions.} 
Students attempted to remove the \textit{complex nested if statements} and \textit{duplicated statement} issues in 23 occasions, but did not always succeed -- the former issue was fixed by two students, the latter issue was fixed by eight students. When working on Exercise 1 to remove the \textit{duplicated statement} issue, seven students reasoned about the presence of unnecessary code. Although this reasoning is adequate, it was followed by incorrect or undesired refactoring steps in three occasions. For example, a student correctly extracted the duplicated statement from the if-else block, but did not move the remaining statement from the else part to the empty if part. Regarding the \textit{complex nested if statements} issue in Exercise 4, four students attempted to merge the nested if statements with an {\&\&}-operator, while ignoring the else part. These steps are instances of refactoring misconceptions \cite{oliveira2023student}: the former one relates to \textit{BadIfElseSimplification}, a misconception when simplifying boolean expressions in if-else blocks, whereas the latter relates to \textit{MergedIfs}, a misconception involving rewriting nested conditional statements. These misconceptions may be caused by an insufficient understanding of applying De Morgan's rules in boolean expressions. This issue was identified in \citeauthor{herman2012describing}'s \cite{herman2012describing} study, and relates to a finding from \citeauthor{da2023test}'s \cite{da2023test} study which investigated student reasoning in code quality. Their results showed that, when explaining code, students occasionally demonstrate an incomplete understanding of loops and conditionals.

\vspace{0.1cm}


\vspace{0.1cm}

\textbf{Threats and limitations.} One limitation of the think-aloud method is the reliance on participants' verbalization. Students may not have verbalized their thoughts all the time. To mitigate this, the first author took note of unclear student actions and verbalizations during the think-aloud session, and asked for clarification in the wrap-up session. Still, we might have missed some thoughts. Another aspect to be considered is that students were observed and recorded throughout the think-aloud activity. This may have affected their behavior, as they may have felt compelled to perform well. In addition, the study was carried out with a small set of code refactoring exercises, which contain a limited set of code quality issues. These exercises were attempted by students from the same university, who took the same programming courses. They participated in the experiment before the lecture on code refactoring, so their prior experience with refactoring was an important factor and may have affected our results. Given the qualitative nature and design of this study, our sample consisted of 12 students, which makes it hard to generalize our results. Further research is needed to improve the validity of our results.

\section{Conclusions and Future Work}
\label{Conclusions}

In a prior study, we investigated incorrect steps students take when refactoring code, and identified a set of misconceptions associated with such steps. To help students resolve these refactoring misconceptions, it is crucial to first get more insight into their reasonings while refactoring code. Therefore, in the present work we conducted a think-aloud study to investigate what CS students think when working on code refactoring exercises. This paper makes the following contributions: a description of students' reasonings while refactoring method-level code, and an analysis of the relation between the code quality issues addressed by students and their reasonings in code refactoring. 

We identified eight categories of student reasoning while refactoring code, which are either related to the presence of code quality issues, the improvement of code quality attributes, or code semantics. We observed a relation between programming experience and reasoning about quality attributes. More experienced students reason more often about quality attributes of code that can be improved, such as readability, rather than look for code quality issues. A possible explanation for this is that experienced students are more familiar with maintaining code from other developers, as well as having their code read by others.

Our results show that students addressed most code quality issues in the refactoring exercises, but often overlooked the presence of a method with multiple responsibilities or the use of a less suitable loop structure. In addition, a few refactoring misconceptions identified in our prior study appeared again, namely a misconception related to combining nested if statements and a misconception related to simplifying an if-else statement. Students often mentioned the presence of unnecessary code as a reason to refactor these conditional statements. Although this reasoning is adequate, some students took incorrect refactoring steps that resulted in semantically incorrect code. We think that these errors may be related to an insufficient understanding of applying De Morgan's rules in boolean expressions.


In future work, we plan to improve an existing refactoring tutoring system to offer adequate feedback on particular refactoring steps. At a later stage, we plan to investigate the effects of using this tutoring system on students' learning of code refactoring.



\balance
\bibliographystyle{ACM-Reference-Format}
\bibliography{bib/bibliography}

\end{document}